\documentstyle[12pt,epsf,epsfig]{article}
\newcommand{\be}{\begin{equation}}
\newcommand{\eea}{\end{eqnarray}}
\newcommand{\beq}{\begin{equation}}
\newcommand{\eeq}{\end{equation}}
\newcommand{\bea}{\begin{eqnarray}}
\newcommand{\ee}{\end{equation}}

\def\Journal#1#2#3#4{{#1} {\bf #2} (#4) #3}

\def\NPB{\em Nucl. Phys.}
\def\PLB{\em Phys. Lett.}

\def\CMP{\em Commun. Math. Phys.}

\begin{document}
\pagestyle{empty} 
\begin{flushright}
 {ROMA-1322/01} 
\end{flushright}
\vskip 1.5cm
\centerline{\LARGE{\bf{KAON PHYSICS FROM LATTICE QCD }}}
\vskip 2.3cm
\centerline{\bf{G. Martinelli}}
\vskip 1.5cm
\centerline{Dip. di Fisica, Univ. ``La Sapienza"  and}
\centerline{INFN, Sezione di Roma, P.le A. Moro, I-00185 Rome, Italy.}
\vskip 4cm
\abstract{
The latest lattice results on  kaon decays and mixing
 are reviewed.  The discussion is focused on recent theoretical progress and new numerical calculations 
 which appeared after Kaon 1999. }
\vfill\eject
\pagestyle{empty}
\clearpage
\setcounter{page}{1}
\pagestyle{plain}
\newpage
\pagestyle{plain} \setcounter{page}{1}
\section{Introduction}
\label{sec:intro}
In this talk  a review  of theoretical and numerical results 
on kaon decays,  obtained with lattice QCD after   Kaon 
1999, is presented.    Several issues of interest for the 
lattice approach can also be found  in the talk  by J. Donoghue at 
this Conference, to which I will refer when necessary.   This
contribution is being written  just  after the  Lattice 2001 Conference~\cite{latt2001}  where new intriguing 
results for $\epsilon^{\prime}/\epsilon$ by 
the CP-PACS~\cite{cppacs2001} and RBC~\cite{rbc2001} Collaborations have 
been presented. I added a discussion of these 
results to the material presented at Kaon 2001.   The main topics 
in this review  are:  i) the determination of  the $\Delta I=1/2$ $K \to 
\pi\pi$ amplitudes; 
ii) the main contributions to  $\epsilon^{\prime}/\epsilon$ due 
to the electropenguin operators ($Q_{7}$ and $Q_{8}$ in the standard 
notation) and to the strong penguin operator $Q_{6}$. 
The most important novelties since Kaon 1999 are the following: 
\vskip 0.1 cm {\bf Theoretical advances}
\begin{itemize}
\item  Detailed and extended studies of the chiral expansion of 
several  relevant amplitudes  in finite and infinite volumes, in the 
quenched and unquenched cases, have been done  in 
refs.~\cite{bernardgolterman}--\cite{goltepallalast}.  Further one-loop 
chiral calculations 
which are particularly important to control   finite volume 
corrections and quenching effects in actual numerical simulations  remain 
to be done. These calculations will also be very useful 
to extract the value of the physical $K \to \pi\pi$ amplitudes from the lattice $K \to \pi$  matrix 
elements  and to extrapolate both  $K \to \pi$ and   $K \to \pi\pi$  
matrix elements, which  are at present computed  at unphysical values of the 
quark masses, to the physical point; 
\item The importance of Final State Interaction (FSI) effects 
for  $\Delta I=1/2$ transitions was first noticed in 
refs.~\cite{truong,isgur}. More recently it has  been 
emphasized  by   Bertolini, 
Eeg and Fabbrichesi~\cite{bertolini} and  Pallante and 
Pich~\cite{PP} that they may have large effects also for $\epsilon^{\prime}/\epsilon$.
 Although the quantitative evaluation of  FSI effects  is still 
controversial~\cite{burasrome}, there is a general 
consensus that,  for a reliable calculation of kaon decay amplitudes,
it is necessary to have a good theoretical control of 
FSI. This  problem  poses serious difficulties  to lattice calculations based on the 
extraction of the $K \to \pi\pi$ amplitudes from the $K \to \pi$ 
matrix elements using chiral perturbation theory ($\chi$PT), as done by the 
CP-PACS~\cite{cppacs2001} and RBC Collaborations~\cite{rbc2001};
\item  The Maiani-Testa no-go theorem ~\cite{mt} has prevented for a long time 
any attempt to directly calculate  $K \to \pi\pi$ matrix elements in an  Euclidean 
lattice.  An important step towards the solution of this
problem has recently been achieved by Lellouch and
L\"uscher in~ref.~\cite{lll} (denoted in the following by LL), who
derived a relation between the $K\to\pi\pi$ matrix elements in a
finite Euclidean volume and the physical kaon-decay amplitudes. 
Further investigation in this direction has been recently done in 
ref.~\cite{lmst} and exploratory numerical studies are just beginning. A 
discussion of these papers will be presented in the following.
\end{itemize}
\vskip 0.1 cm {\bf New numerical results}
\begin{itemize}
\item A lattice calculation of matrix element of the electro-magnetic operator  
$\langle  \pi^{0} \vert \hat Q^{+}_{\gamma} \vert K^{0} \rangle$,  
which is relevant for the CP violating $K_{L} \to 
\pi^{0}e^{+}e^{-}$ decay,  has been performed in ref.~\cite{damir1}; 
\item New results for $K^{0}$--$\bar K^{0}$ mixing with Improved 
Wilson Fermions~\cite{Becirevic:2001ki,damir2} and Domain Wall Fermions 
(DWF)~\cite{rbc2000,cppacsk0} have appeared.  With 
some differences,  they 
confirm previous lattice results~\cite{lellouch2000}. Calculations of
 $\langle \bar  K^{0} \vert \hat Q_{i} \vert K^{0} \rangle$  for all 
possible  operators $Q_{i}$ which can mediate $K^{0}$--$\bar K^{0}$ mixing in extensions of the 
Standard Model have  be done in   refs.~\cite{damir2,doninids2}. These results 
are very useful  to put severe constraints on FCNC parameters of SUSY 
models~\cite{footballteam};
\item The   SPQ$_{CD}$R 
Collaboration has presented results for the amplitudes
$\langle \pi^{+} \pi^{0} \vert \hat Q_{4}   \vert K^{+} \rangle$ 
and   $\langle \pi \pi \vert \hat Q_{7,8}   \vert K^{0}\rangle_{I=2}$, obtained by following the approach of~\cite{lll} and \cite{lmst}  of  
computing directly the $K \to \pi\pi$ amplitude~\cite{dlin2001,spqr}.  
 The chiral 
behaviour of the matrix elements has also been studied in view of the 
extrapolation to the physical point and the necessary calculations of 
the relevant chiral loops are  under way;
\item  In a first exploratory study, the  SPQ$_{CD}$R Collaboration has also observed the first 
signal for  $\langle \pi \pi \vert \hat Q^{-}   \vert K^{0} 
\rangle_{I=0}$  and $\langle \pi \pi \vert \hat Q_{6}  \vert K^{0} 
\rangle_{I=0}$~\cite{spqr};
\item As mentioned before,  CP-PACS and RBC have presented results 
for the CP conserving $\Delta I=1/2$ amplitude, ${\cal A}_{0}$, and 
for $\epsilon^{\prime}/\epsilon$,
obtained from $K \to  \pi$ matrix elements using soft pion theorems. 
Both groups find a negative value, 
$\epsilon^{\prime}/\epsilon\sim -5 \times 10^{-4}$. The 
possible origin of the discrepancy with the experimental 
results~\cite{na48,ktev} will be discussed in detail.
\end{itemize}
In this short review a severe selection of arguments has been necessary. 
The reader,  interested to have more information about lattice 
calculations of weak decays,  including heavy flavours, may address 
either the nice reviews by L.~Lellouch~\cite{lellouch2000}  and C.~Bernard~\cite{bernard2000}  
 at Lattice 2000, or the fortcoming 
Proceedings of the 2001 Lattice Conference~\cite{latt2001}.
\section{General framework}
Physical kaon weak decay amplitudes can be described with 
negligible error  (of ${\cal O}(\mu^{2}/M_{W}^{2})$, where $\mu$ is 
the renormalization scale) in terms of   matrix elements of the effective 
weak Hamiltonian 
\beq  \langle \pi \pi \vert {\cal H}^{\Delta S=1} \vert K 
\rangle \, , \eeq
 written as combination of  
Wilson coefficients and renormalized local  operators 
\beq {\cal H}^{\Delta S=1}= -\frac{G_{F}}{\sqrt{2}} \sum_{i}  C_{i}(\mu) 
\hat Q_{i}(\mu) \,  .  \eeq 
The sum is over a complete set of operators, which depend on  $\mu$. In general there are 12 four-fermion operators 
and  two  dimension-five  operators: a  chromomagnetic one and an 
electromagnetic one. In the Standard Model, the contribution of the dimension-five operators 
is usually neglected (SUSY effects may enhance the  contribution of the chromomagnetic 
operator~\cite{murayama}).   The calculation of the matrix elements 
must be done non-perturbatively, and this is the r\^ole of the 
lattice, whereas the Wilson coefficients can be computed 
in perturbation theory.
\subsection*{Wilson coefficients and renormalized operators}
For ${\cal H}^{\Delta S=1}$  the Wilson 
coefficients have been computed at the next-to-leading order in 
refs.~\cite{alta}--\cite{ciuc2}.  The perturbative calculation  is reliable  provided that the scale $\mu$ is large enough, $\mu 
\gg \Lambda_{QCD}$. In this 
respect calculations performed below the charm quark mass ($m_{c} \sim 
1.3$~GeV)  are, in my opinion, suspicious. In fact, either $\mu \ll 
m_{c}$, and then perturbation theory is questionable, or $\mu \sim 
m_{c}$, and then the effective three-flavour weak Hamiltonian (with 
propagating up, down and strange quarks) cannot 
be properly matched to the four-flavour theory (up,down, strange and charm) 
because of the presence of  operators of higher dimension which 
contribute at ${\cal O}(\mu^{2}/m_{c}^{2})$.  
Another  important remark is in order.  Wilson coefficients and   matrix elements of the  operators 
$\hat Q_{i}(\mu)$ separately
depend on the choice of the renormalization scale and scheme. This dependence  cancel in
physical quantities, up to higher-order corrections in the perturbative expansion of
the Wilson coefficients.  For this crucial cancellation to take  place, the non-perturbative method 
used to compute the hadronic matrix elements must allow   a definition of the
renormalized operators consistent with the scheme used in the 
calculation of the Wilson coefficients.   The lattice approach 
satisfies this requirement. 

Matching of  bare (divergent) lattice 
operators, $Q_{i}(a)$ to the renormalized ones is obtained  by computing suitable 
renormalization constants $Z_{ik}(\mu a)$
\be {\cal A}^{i}_{I=0,2}(\mu) = \langle \pi \pi \vert \hat Q_{i}(\mu)  \vert K
\rangle_{I=0,2} = \sum_{k} \, Z_{ik}(\mu a) \langle \pi \pi \vert  Q_{k}(a)  \vert K
\rangle_{I=0,2} \, , \label{eq:renor}\ee
where $a$ is the lattice spacing.
The  \textit{ultra-violet} (UV) problem,
which deals with the construction of finite matrix elements of
renormalized operators constructed from the bare lattice ones, has
been addressed in a series of papers~\cite{wise}-\cite{capitani} 
and is, at least in principle, completely solved. The remaining 
difficulties are practical ones. On the one hand, the $\Delta I=1/2$ 
operators suffer from  power divergences in the ultraviolet cutoff, 
$1/a$.  These divergences, that cannot be subtracted using  perturbation 
theory~\cite{lellouch2000},  can be eliminated by performing suitable 
numerical  subtractions. The subtraction procedure, however, suffers from systematic uncertainties which are 
difficult to keep under control,  see below.  On the other hand, the  
perturbative calculation of   the logarithmically divergent (or 
finite)    $Z_{ik}(\mu a)$   is rather inaccurate. Several non-perturbative methods have been 
developed in order to compute $Z_{ik}(\mu a)$   
non-perturbatively~\cite{doninids2}, \cite{npm}\--\cite{lleshouches} and the uncertainties 
vary between  1\%, for the simplest bilinear operators, to $10\div 25$\%, in the case of the 
four-fermion operators of interest. An accurate determination of 
$Z_{ik}(\mu a)$  for the $\Delta I = 1/2$ operators is missing to date 
and more work is needed in this direction. 

Since numerical simulations 
are performed at finite values of the  lattice spacing, $a^{-1} \sim 
2\div 4$~GeV, another source of uncertainty in the determination of the matrix 
elements  comes from discretization errors. They  are of ${\cal O}(\Lambda_{QCD} a)$,  ${\cal  O}(\vert \vec p\vert a)$ or ${\cal  
O}(m_{q} a)$, where $\vec p$ is a typical  hadron momentum and $m_{q}$ the quark mass.  The  simplest strategy to reduce discretization effects  
consists in   computing physical  quantities at several values of the lattice spacing and then 
extrapolate to $a \to 0$.  A different  approach, pioneered by 
Symanzik  and extensively studied on the 
lattice~\cite{lleshouches}, is to reduce discretization errors from 
${\cal O}(a)$ to ${\cal O}(a^{2})$ by improving the lattice action 
and operators. This method, which can also be combined with the 
extrapolation to $a\to 0$, has been succesfully applied  to the 
determination of the quark masses and of matrix elements of bilinear 
operators. A systematic study of the improvement of four fermion 
operators is still to be done. Note that with DWF~\cite{dwf} or overlapping 
Fermions~\cite{of} the errors are automatically of  ${\cal 
O}(a^{2})$~\cite{pilar}. These 
formulations of lattice QCD are, however, much more demanding in terms 
of computing resources.  Discretization errors correspond to the 
matching problems in effective theories with a low cutoff recently 
discussed  by Cirigliano, Donoghue and Golowich~\cite{cdg}. In this 
respect the problem is softer for the lattice approach since numerical 
simulations are already performed at relatively large scales.
\subsection*{$K\to \pi$ and $K \to \pi\pi$ matrix elements}
Two main roads have been suggested in the past in order to obtain ${\cal 
A}^{i}_{I=0,2}(\mu)$:
\begin{itemize} 
\item Compute the $K \to 0$ and $K \to \pi$ matrix elements $\langle 
0 \vert \hat Q_{i}(\mu)  \vert K \rangle$ and $\langle \pi \vert \hat Q_{i}(\mu)  \vert K
\rangle$  and then derive $\langle \pi \pi \vert \hat Q_{i}(\mu)  \vert K
\rangle_{I=0,2}$  using soft pion 
theorems~\cite{wise,renorm}. In this case the $K \to 
\pi\pi$ amplitudes can be evaluated only at the lowest order of the 
chiral expansion;
\item Compute directly $\langle \pi \pi \vert \hat Q_{i}(\mu)  \vert K
\rangle_{I=0,2}$~\cite{lll,lmst,draper,dawson}.
\end{itemize} 
The main difficulty in the latter case  is due to 
the relation between $K \to \pi\pi$ matrix elements computed in a
finite Euclidean space-time volume and the corresponding physical
amplitudes (the \textit{infrared} problem).  In the approach where the 
$A^{i}_{I=0,2}(\mu)$ are  extracted from the one-pion matrix elements,
$\langle \pi  \vert \hat Q_{i}(\mu)  \vert K \rangle$, the problem 
is the size of the corrections which relate matrix elements  in 
the chiral limit to the corresponding  physical amplitudes. These corrections are 
expected to be much larger for  $I=0$ amplitudes, because of the larger 
FSI~\cite{PP}. Both the approaches have their advantages  and drawbacks to which 
most of the following discussion will be devoted. 

The infrared problem arises from two sources:
i) the unavoidable continuation of the theory to
Euclidean space-time and
ii) the  use of a finite volume in numerical simulations.
An important progress has  been achieved by LL, who
derived a relation between the lattice $K\to\pi\pi$ matrix elements in a
finite volume and the physical kaon-decay amplitudes~\cite{lll}.
An alternative discussion of boundary
effects and the LL-formula, based on a study of the properties of
correlators of local operators was presented in ref.~\cite{lmst}. In this approach the LL-formula   
is derived for all elastic states under the inelastic threshold, with exponential
accuracy in the quantization volume. As a consequence  the relation between finite-volume matrix
elements and physical amplitudes, derived by Lellouch and
L\"uscher for the lowest seven energy eigenstates, can be extended
to all elastic states under the inelastic threshold. It can also been
explicitly  demonstrated how finite volume correlators
converge to the corresponding ones in infinite volume.
The derivation of~\cite{lmst}  is based on the property of correlators of local
operators which can be expressed, with exponential accuracy, both
as a sum or as an integral over intermediate states, when
considering volumes larger than the interaction radius and
Euclidean times $0<t \simeq L$.  It can also been shown that it is possible to extract  $K \to \pi\pi$
amplitudes also when the kaon mass, $m_K$, does not match the
two-pion energy, namely when the inserted operator carries a
non-zero energy-momentum. Such amplitudes  are very useful, for
example, in the determination of the coefficients of  operators
appearing at higher orders in $\chi$PT, as illustrated by the numerical 
results for  $\Delta I=3/2$ transitions presented in sec.~\ref{sec:numerical}.

Let us sketch now the derivation  of the result with an illustrative 
example which is not explicitly written  in~\cite{lmst}.
Consider the following Euclidean T-products (correlators):
\bea G(t,t_{K}) &=& \langle 0 \vert T[J(t) \hat Q_{i}(0)   
K^{\dagger}(t_{K})]\vert  0\rangle \, ,  \nonumber \\
G(t) &=& \langle 0 \vert T[J(t) J(0)]   \vert  0\rangle \, ,\quad \quad 
G(t_{K}) = \langle 0 \vert T[ K(t_{K}) K^{\dagger}(0)\vert  0\rangle 
\, , 
\label{eq:corres} \eea
where $J$ is a scalar operator which excites (annhilates) zero angular 
momentum $\pi\pi$ states from (to) the vacuum and $K^{\dagger}$ is a pseudoscalar source which excites a kaon from the vacuum 
($t > 0$ and  $t_{K}  < 0$). At large time distances we have 
\bea G(t,t_{K}) &\to&  V \, \sum_{n} \langle 0 \vert J \vert  
\pi\pi(n)\rangle_{V} \langle  \pi\pi(n) \vert \hat Q_{i} \vert  K\rangle_{V}  \langle K \vert    K^{\dagger} \vert  0\rangle_{V} 
\, \exp(-W_{n} \, t -m_{K}\, t_{K}) \, , \nonumber \\
G(t) &\to&  V \, \sum_{n} \langle 0 \vert J \vert  
\pi\pi(n)\rangle_{V} \langle  \pi\pi(n) \vert J \vert 0\rangle_{V}  
\, \exp(-W_{n} \, t ) \, , \nonumber \\
G(t_{K}) &\to&  V  \langle 0 \vert K \vert  
K \rangle_{V} \langle  K \vert K^{\dagger}\vert 0\rangle_{V}  
\, \exp( -m_{K}\, t_{K} )\, . \eea
From a study of the time dependence of $G(t,t_{K})$, $G(t)$ and 
$G(t_{K})$ we may extract  \begin{enumerate} \item the kaon mass 
$m_{K}$; \item the two-pion energies on the finite lattice volume, 
$W_{n}$. As shown by M.~L\"uscher in~\cite{ml}, the $W_{n}$ are related 
to the infinite volume phase-shift of the 
two pions, $\delta(k)$,  via the following relations 
\bea
W_{n} = 2 \, \sqrt{m_{\pi}^{2} + k^{2} }  \, , \quad \quad 
\frac {\phi (q) + \delta(k)}{\pi} 
 = n \label{phase} \, , \quad n=1,2, \dots \, ,
\quad  q & = &\frac{k L}{2\pi}  \, , \label{eq:wn}
\eea where $n$ is a non-negative integer~\footnote{ For $n=0$, there
are two solutions: one corresponding to $k=0$ which is spurious,
the other giving the L\"uscher relation between the finite volume
energy and the  scattering length.} and  the function $\phi(q)$ is defined in \cite{ml}.
\item the operator matrix elements on a finite volume 
$\langle  \pi\pi(n) \vert \hat Q_{i} \vert  K\rangle_{V}$, $\langle  
\pi\pi(n) \vert J \vert 0\rangle_{V}$ and $\langle K \vert K^{\dagger}\vert 
0\rangle_{V} $. \end{enumerate}
Moreover by a suitable choice of the lattice parameters it is 
possible to match one of the two-pion energies in such a way that 
$m_{K}=W_{n^{*}}$. In practice, since   it will be possible to disentangle only the first few states, 
the matrix elements will be computed with  $n^{*}=0\div 2$.

The fundamental point is that it is possible to relate  the finite-volume Euclidean matrix element 
$\langle  \pi\pi(n^{*}) \vert \hat Q_{i} \vert  K\rangle_{V}$  with 
the absolute value of the physical amplitude  $\langle  \pi\pi \vert \hat 
Q_{i} \vert  K\rangle$:
\bea  \langle  \pi\pi \vert \hat  Q_{i} \vert  K\rangle = 
\sqrt{{\cal F}} \,   \langle \pi\pi(n^{*}) \vert \hat Q_{i} \vert  K\rangle_{V} 
\, , \quad \quad 
{\cal F}=32 \pi^{2} V^2 \, \frac{\rho_{V}(E) E m_K }{\kappa(E)}\, , 
\label{eq:rela}\eea
where
\beq  \kappa(E)=  \sqrt{\frac{E^{2}}{4}-m_{\pi}^{2}} \, , 
\quad \quad \rho _{V}(E)=\frac{dn}{dE} = \frac{q\phi^\prime(q)
+k\delta^\prime(k)}{4 \pi k^2}E\, . \label{density} \eeq
The last expression in
eq.~(\ref{density}) is the one  which one would heuristically
derive from a na\"{\i}ve interpretation of $\rho_V(E)$ as the
density of states, cf. eq.~(\ref{phase}). There are however, some
technical subtleties with such an interpretation which will not be
discussed here. Eq.~(\ref{eq:rela}) holds also when  $m_{K} \neq W_{n}$ and the 
operator carries non-zero energy-momentum~\cite{lmst}.  

By varying the kaon and 
pion masses and momenta, one may  then fit the coefficients of the chiral 
expansion  of $\langle  \pi\pi \vert \hat  Q_{i} \vert  K\rangle$  and 
use these  coefficients to extrapolate to the physical point. This procedure 
is necessary, at  present, since it is not possible  to 
compute the matrix elements at realistic  values of the quark masses and  in the unquenched case. To give an explicit example, the matrix 
elements of $\hat Q_{7,8}$ (for generic meson masses, one pion at rest 
and the other with a given  momentum corresponding to an energy 
$E_{\pi}$)  can be written as~\cite{dlin2001,cirigliano} 
\bea  &&-i {\cal M}_{7,8} (m_{K}, m_{\pi}, E_\pi) =
\gamma^{7,8}  + \delta^{7,8}_{1}\, \left( \frac{ m_K \, 
(m_\pi+E_\pi)}{2} - m_\pi E_\pi  \right)
  \nonumber \\ && +  \delta^{7,8}_2 \, \left( -\frac{ m_K \, ( m_\pi+E_\pi)}{2} - 
 m_\pi   
 E_\pi  \right)    - \delta^{7,8}_3  m_K \, ( m_\pi+E_\pi) +  \left(  \delta^{7,8}_4  + \delta^{7,8}_5 
\right) \left( 2  m^2_K + 4 m^2_\pi \right) \nonumber \\ &&
   + \delta^{7,8}_6 \left( 4  m^2_K + 2 m^2_\pi \right) 
 +\mbox{chiral}  \ \  \mbox{logs} \ \ + {\cal O}(p^{4})  \label{eq:eo78} 
 \\
 \, .  \nonumber  \eea
A fit to the lattice data for ${\cal M}_{7,8}$, extracted from 
suitable correlation functions,  allows  the determination of the 
couplings $\gamma^{7,8}$, $\delta^{7,8}_{1}$, etc. The results of the 
extrapolated amplitudes, if 
chiral logarithms are included, are independent of the cutoff used  in the chiral theory.
\begin{figure}[htb!]
\begin{center}
{\epsfig{figure=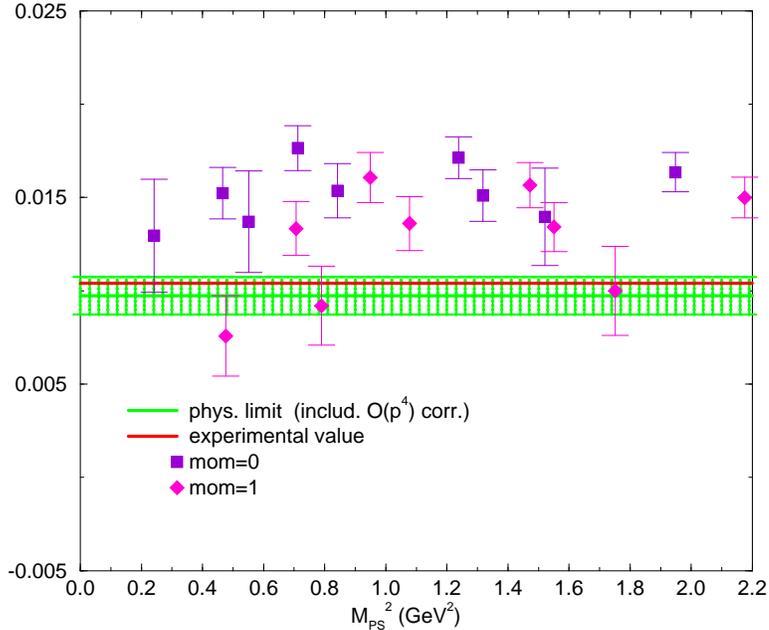,angle=270,width=0.7\linewidth}}
\end{center}
\caption{{\it Chiral behaviour of the matrix element of $Q_{4}$,  ${\cal 
M}_{4} (m_{K}, m_{\pi}, 
 E_\pi)$. The extrapolation to the physical point using only the data 
corresponding to the lightest masses, without the effect of the chiral 
logarithms and using the operators of ref.~\cite{gopa}, is also shown 
as a shadowed band. The experimental result, computed as explained in 
the text,  is also given as a line.}}
\label{fig:q4}
\end{figure}
\begin{figure}[htb!]
\begin{center}
{\epsfig{figure=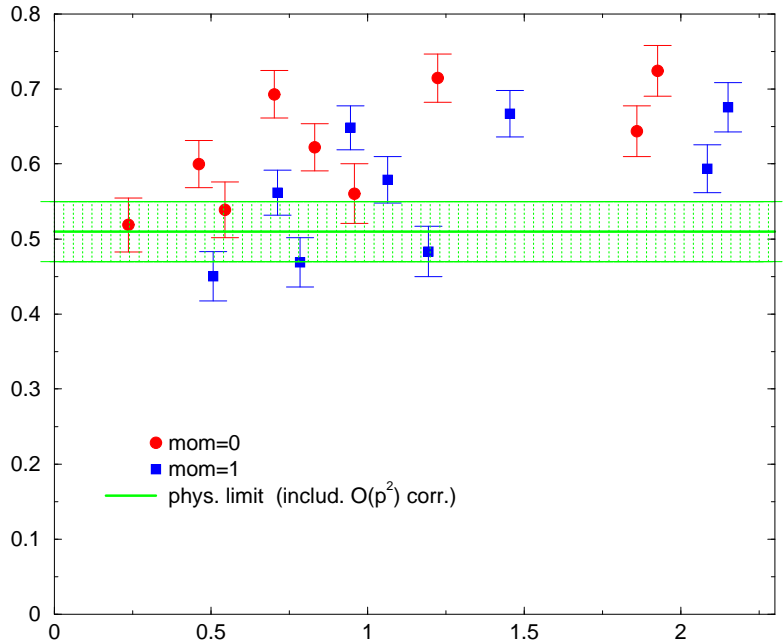,angle=270,width=0.6\linewidth}}
\end{center}
\caption{{\it Chiral behaviour of the matrix element of $Q_{8}$,  ${\cal M}_{8} (m_{K}, m_{\pi}, 
 E_\pi)$. The 
extrapolation to the physical point using only the data 
corresponding to the lightest masses, without the effect of the chiral 
logarithms and using the operators of ref.~\cite{cirigliano}, is also shown 
 as a shadowed band. }}
\label{fig:q8}
\end{figure}

\section{Numerical Results} 
\label{sec:numerical}.

In this section, the latest numerical results from lattice QCD for $K\to \pi \pi$  and $K\to \pi$  
amplitudes and for $\hat B_{K}$  are  reviewed, together with a 
comparison  with other approaches.  
\subsection*{$\Delta I=3/2$ transitions and $K^{0}$--$\bar K^{0}$ 
mixing}
The SPQ$_{CD}$R Collaboration~\cite{dlin2001,spqr} has presented (quenched) results for 
both $\Delta I=3/2$ and $\Delta I=1/2$ amplitudes, obtained  from a calculation of  $K \to \pi\pi$ matrix elements
following  the strategy of refs.~\cite{lll,lmst}. 
For  $\Delta I=3/2$ transitions,   an extended study of the 
chiral behaviour of the matrix elements of the operators $\hat Q_{4,7,8}$, 
renormalized non-perturbatively, was performed.  In 
figs.~\ref{fig:q4}  and ~\ref{fig:q8},  ${\cal M}_{4} (m_{K}, m_{\pi} 
E_\pi)$   and ${\cal M}_{8} (m_{K}, m_{\pi} 
E_\pi)$   are  shown as a function of the kaon mass. Note that the 
amplitude does  depend on three independent quantities that are let to 
vary, namely $m_{K}$, $m_{\pi}$ and $E_{\pi}$ (one of the two pions is always 
at rest).  In the same figures  the extrapolation to the physical 
point,  performed by  using the chiral expansion of the matrix elements  as in 
eq.~(\ref{eq:eo78}), is given as a band. In  fig.~\ref{fig:q4}  the 
experimental number, extracted from the $K^{+}  \to \pi^{+} \pi^{0}$ 
decay rate   using the Wilson coefficient of $\hat Q_{4}$ computed at the 
NLO, is also shown. 
The extrapolations are preliminary, since they do not include the 
effects of  (quenched) chiral logarithms that  have not been 
computed yet for the kinematical configurations used in this study.   
For some of the points,  masses and momenta are too large to 
use chiral perturbation theory, and they have not been included in 
the fit.
The  preliminary results in  fig.~\ref{fig:q4} already  give us an interesting 
physics information. In the chiral limit, and using $SU(3)$ symmetry,  one may relate the $K^{+}  \to 
\pi^{+} \pi^{0}$  amplitude to the $K^{0}$--$\bar K^{0}$ mixing  parameter $\hat B_{K}$~\cite{olddono}.  In this limit, a large 
value for $\hat B_{K}$,  as found in lattice calculations and  unitarity 
triangle analyses~\cite{ciu2001}, $\hat  B_{K}\sim 0.85$, would lead 
to a value of the  $K^{+}  \to  \pi^{+} \pi^{0}$  amplitude   larger than 
the experimental one by $\sim  50\%$.   The extrapolation of the 
results of fig.~\ref{fig:q4}, together with those for $\hat B_{K}$ 
obtained by the same collaboration~\cite{damir2}  given in 
table~\ref{tab:bks}, demonstrate that chiral corrections to both the  $K^{+}  \to  \pi^{+} \pi^{0}$  amplitude 
and $\hat B_{K}$  can easily reconciliate a large value of $\hat B_{K}$
with the experimental $K^{+}  \to \pi^{+} \pi^{0}$ amplitude.  

It is very instructive to compare the results for $\hat Q_{7}$ and 
$\hat Q_{8}$ 
with those obtained in other approaches, table~\ref{tab:q78}. For the 
sake of comparison all the results are converted to the 
$\overline{MS}$ scheme at a renormalization scale $\mu=2$~GeV. In the 
case of $\hat Q_{8}$ 
we notice the nice agreement between  lattice results obtained 
from $K \to \pi$ matrix elements using chiral perturbation theory and  
those from the first direct calculation of  $K \to \pi\pi$  
amplitudes.  Similar numbers were obtained by the CP-PACS~\cite{cppacs2001} and RBC~\cite{rbc2001} 
collaborations, which computed $K\to \pi$ matrix elements with DWF. 
These groups presented  the results in different renormalization schemes 
and scales and for this reason the values have not been included in 
the table. There is,  however,  a  very large discrepancy of the lattice 
results with those  obtained with dispersive methods~\cite{dongol} or 
with the $1/N_{c}$  expansion~\cite{derafael},  whose results  would correspond to a huge value of the $B$-parameter,  
$B_{8}=3\div 7$~\footnote{ Other results for $\hat Q_{7}$ and $\hat Q_{8}$  can 
be found in refs.~\cite{bertolini} and \cite{hambye}.}. 
In  order to reproduce the experimental results for 
$\epsilon^{\prime}/\epsilon$ within the Standard Model, such a large 
value of $B_{8}$ implies  a stratosferic value for $B_{6}$. After Kaon 
1999 a new analysis,  performed with spectral function techniques by Bijnens, Gamiz 
and Prades appeared~\cite{Bijnens:2001ps}. In this paper, a value of $\langle 
\hat  Q_{8}\rangle$    much lower than in refs.~\cite{dongol,derafael} was found,
although still about a factor of two larger than  lattice determinations. The very low 
value of $\langle \hat   Q_7\rangle$  found from the lattice $K \to \pi\pi$ 
calculation originates from large cancellations occuring in the 
renormalization of the relevant operator and requires further 
investigation.
 \begin{table}[htb]
\centering
\caption{{\it $K \to \pi \pi $ matrix elements in GeV~$^3$, at $\mu=2$~GeV in the 
$\overline{MS}$ scheme, from lattice calculations (first three rows) and other 
approaches.}}
\label{tab:q78}
\begin{tabular}{||c|c|c|c||}\hline\hline
Reference & Method & $\langle \hat   O_8\rangle$ & $\langle  \hat  O_7\rangle$  \\
\hline \hline SPQ$_{CD}$R~\cite{spqr} 2001 & $K \to \pi \pi$ &  
$0.53 \pm 0.06$ &  $0.02 \pm 0.01$ \\ 
SPQ$_{CD}$R~\cite{damir2} 2001 & $K \to \pi$ + $\chi$PT &  
$0.49 \pm 0.06$ &  $0.10 \pm 0.03$ \\
APE~\cite{doninids2} 1999 & $K \to \pi$ + $\chi$PT &  
$0.50 \pm 0.10$ &  $0.11 \pm 0.04$ \\
Amherst~\cite{dongol} 1999 & Dispersion relations &  
$2.22 \pm 0.67$ &  $0.16 \pm 0.10$ \\
Marseille~\cite{derafael} 1998  &$1/N_{c}$ & $3.50 \pm 1.10 $ & $0.11 
\pm 0.03$ \\
BGP~\cite {Bijnens:2001ps} 2001  &Spectral functions & $1.2 \pm 0.5 $ & 
$0.26  \pm 0.03$ \\
\hline \hline
\end{tabular}
\end{table}

Lattice predictions for $\hat B_{K}$ have been very stable over the 
years, with a central value centered around $0.85$. This is 
essentially also the 
value extracted from Unitarity Triangle 
Analyses~\cite{ciu2001}, thus confirming that lattice QCD can 
predict (and not only postdict) physical quantities. In 
table~\ref{tab:bks} the lattice world average is given, together with 
some of the  latest lattice results. $\hat B_{K}$ from CP-PACS~\cite{cppacsbk} 
(with operators renormalized perturbatively)
and RBC~\cite{rbc2000} (with operators renormalized non-perturbatively)  
has been obtained with DWF. This  formulation of QCD  
should garantee a better control of the chiral behaviour of the 
regularized theory with respect to Wilson-like fermions. The 
results from ref.~\cite{damir2} have been obtained using the 
non-perturbatively improved  Wilson-like action, with a new method based on the 
Ward Identities~\cite{nosub}, which save us from the painful subtractions of the 
wrong chirality operators, which was  necessary in the 
past. Similar strategies have been pursued with 
twisted-mass fermions~\cite{sfm}.  The results with DWF are about $15\%$ below 
the world average and show a marked decrease at small quark masses. 
This could reconciliate the large value of $\hat B_{K}$ at the 
physical kaon mass with the low value of this parameter obtained in 
the chiral limit by  ref.~\cite{derafael1}. Whether the decrease at 
small quark masses is a physical effect, or is due to  lattice artefacts 
(finite volume, residual chiral symmetry breaking etc.) 
remains to be investigated.  
 \begin{table}[htb]
\centering
\caption{{\it$B_{K}$ in the 
$\overline{MS}$ scheme at the renormalization scale $\mu=2$~GeV and 
the renormalization group invariant $\hat B_{K}$  from recent lattice calculations. 
The results from SPQ$_{CD}$R, CP-PACS and RBC Collaborations are 
quenched.}}
\label{tab:bks}
\begin{tabular}{||c|c|c|c||}\hline\hline
Reference & Method & $B_{K}^{\overline{MS}} (2\, \mbox{GeV})$ & $\hat B_{K}$  \\
\hline \hline World Average &  & $0.63\pm 0.03\pm 0.10$ &  
$0.86 \pm 0.06\pm 14$  \\ 
SPQ$_{CD}$R~\cite{damir2} 2001 & with subtractions & $0.71\pm 0.13$ & $0.91 
\pm 0.17$ \\
SPQ$_{CD}$R~\cite{damir2} 2001 & Ward identity method & $0.70\pm 0.10$ & 
$0.90  \pm 0.13$ \\
CP-PACS~\cite{cppacsbk} 2001 & DWF Pert. Ren. &    $0.5746(61)(191)$ &  $0.787 \pm 
0.008$ \\
RBC~\cite{rbc2000} 2000  &  DWF Nonpert. Ren.  &   $0.538 \pm 0.08$ &  $0.737 \pm 0.011$ \\
Ciuchini et al.~\cite{ciu2001} 2001 & Unitarity Triangle &  
$0.70^{+0.23}_{-0.11}$  &  $0.90^{+0.30}_{-0.14}$ \\
\hline \hline
\end{tabular}
\end{table}
\subsection*{$\Delta I=1/2$ transitions and  
$\epsilon^{\prime}/\epsilon$}
\begin{figure}[htb!]
\begin{center}
{\epsfig{figure=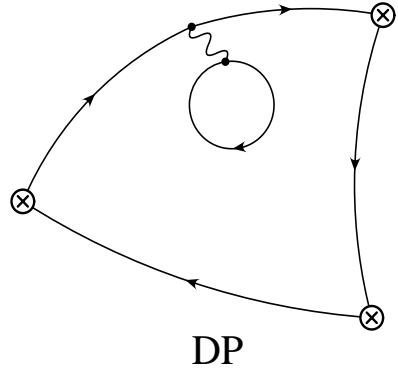,height=0.3\linewidth,width=0.3\linewidth}}
\end{center}
\caption{{\it The disconnected penguin (DP) diagram present in $\Delta 
I=0$ transitions.}}
\label{fig:dp}
\end{figure}

\begin{figure}[htb!]
\begin{center}
{\epsfig{figure=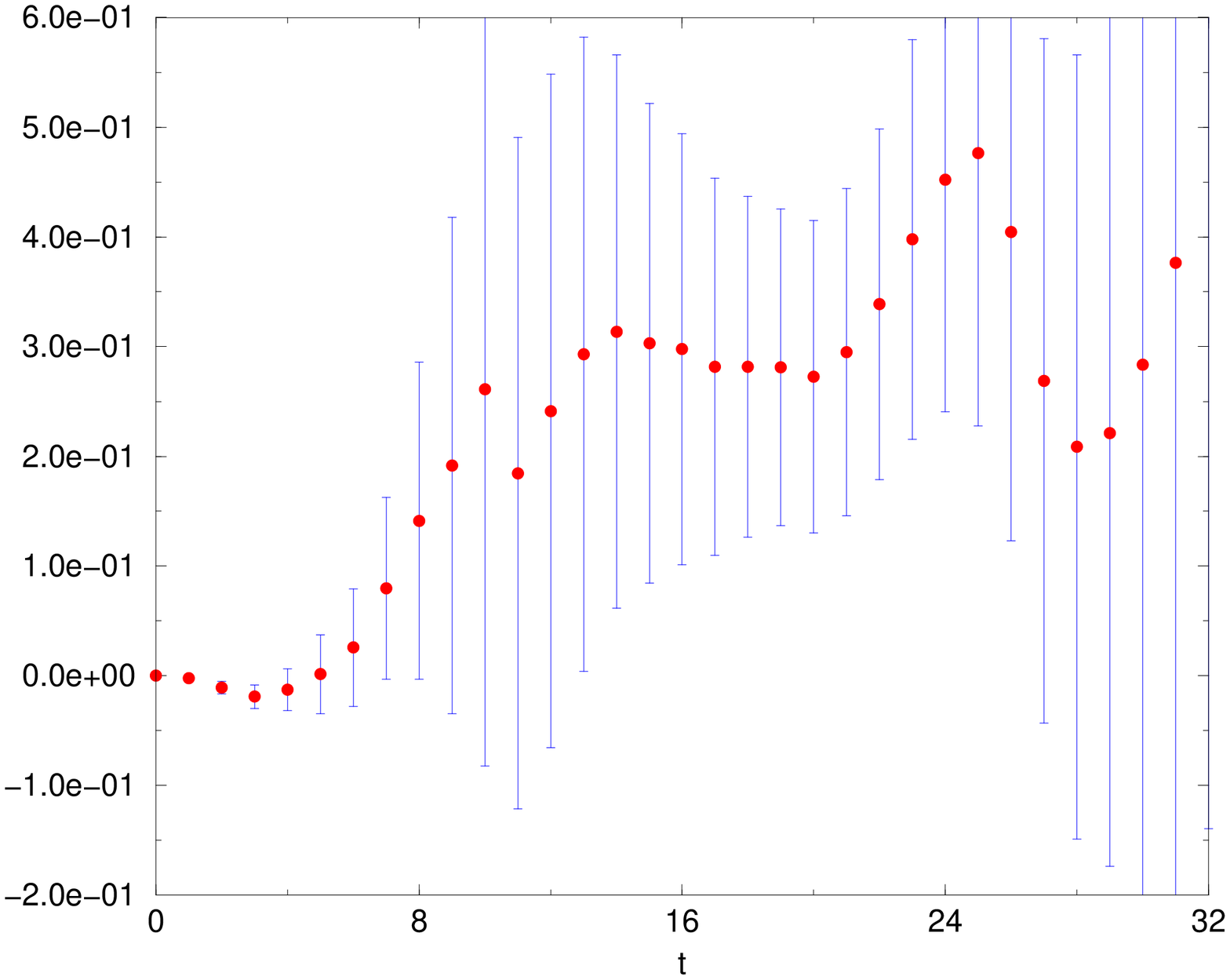,angle=270,width=0.6\linewidth}}
\end{center}
\caption{{\it First signal for the matrix element $\langle \pi \pi 
\vert Q^{-} \vert K \rangle$ at the matching point $m_{K} \sim 
W_{0}$, obtained with 400 configurations  and  $V=24^{3} \times 64$.}}
\label{fig:qm}
\end{figure}

\begin{figure}[htb!]
\begin{center}
{\epsfig{figure=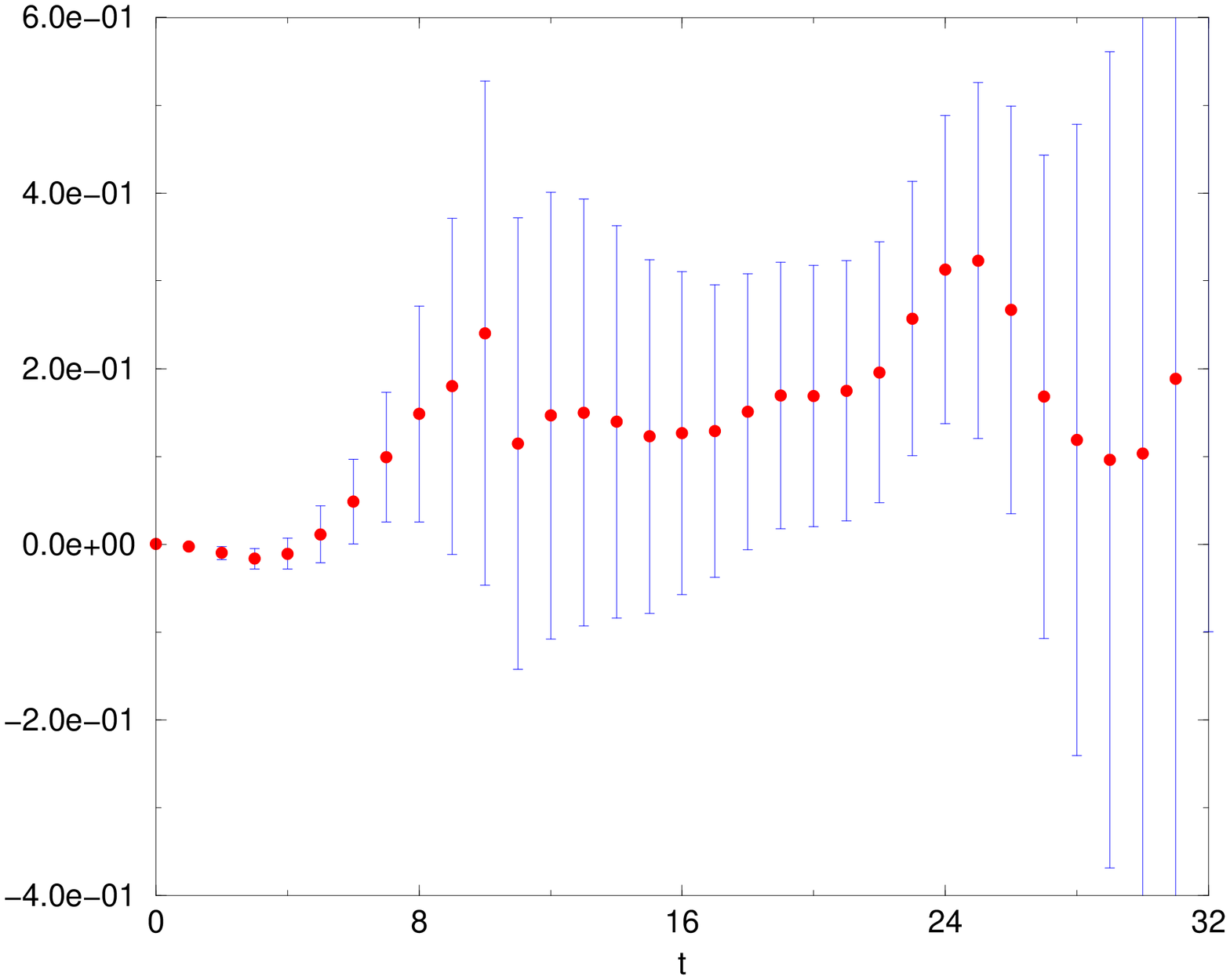,angle=270,width=0.6\linewidth}}
\end{center}
\caption{{\it  First signal for the matrix element $\langle \pi \pi 
\vert Q_{6} \vert K \rangle$ at the matching point $m_{K} \sim 
W_{0}$, obtained with 400 configurations  and  $V=24^{3} \times 64$. }}
\label{fig:q6}
\end{figure}

\begin{figure}[htb!]
\begin{center}
{\epsfig{figure=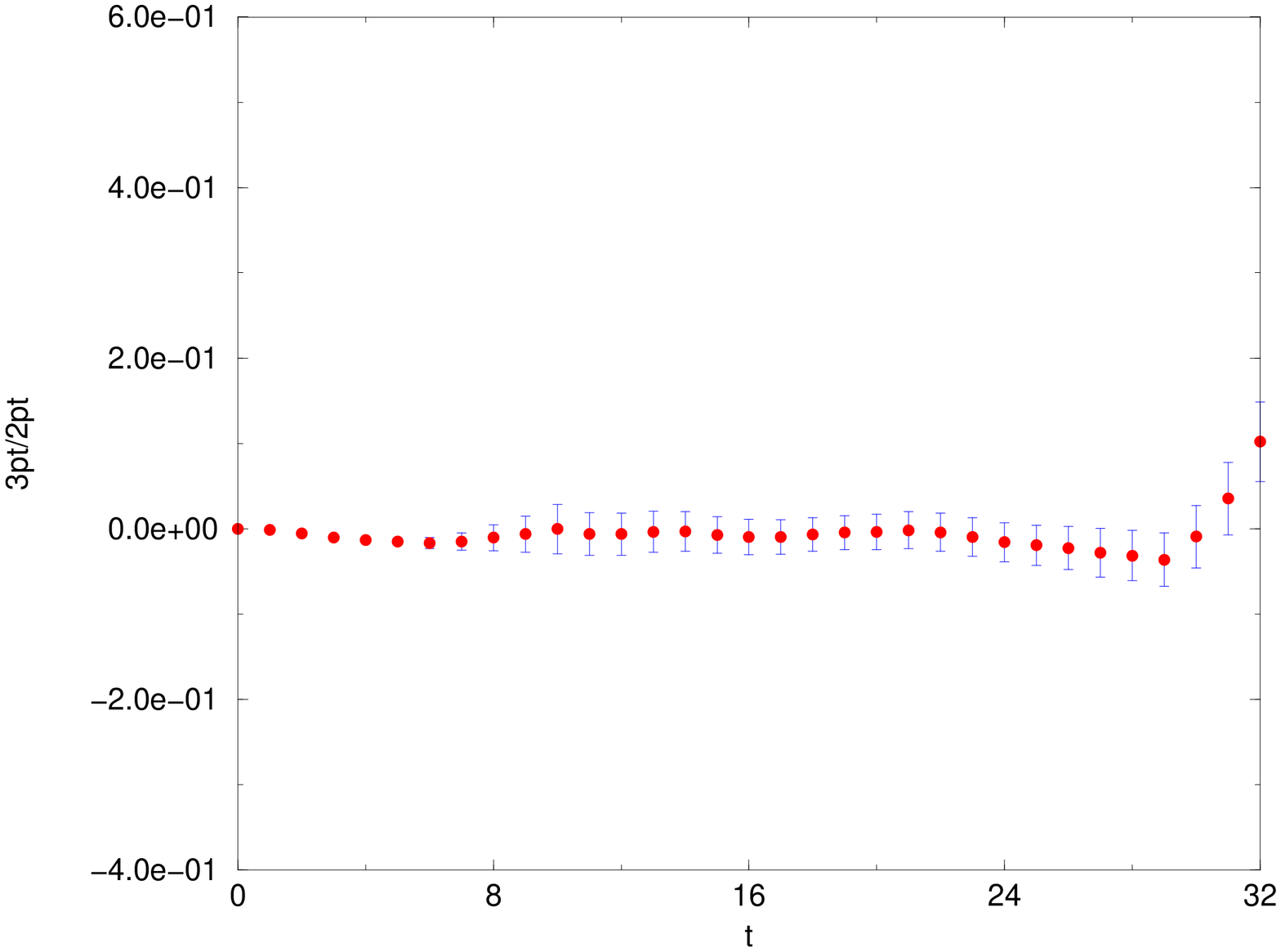,angle=270,width=0.6\linewidth}}
\end{center}
\caption{{\it   $\langle \pi \pi 
\vert \bar s \gamma_{5} d \vert K \rangle$ at the matching point $m_{K} \sim 
W_{0}$, obtained with 400 configurations  and  $V=24^{3} \times 64$. 
At the matching point $\langle \pi \pi 
\vert \bar s \gamma_{5} d \vert K \rangle$ decreases by a huge factor 
and  thus the subtraction for $\langle \pi \pi 
\vert Q_{6} \vert K \rangle$  is very small.}}
\label{fig:qp}
\end{figure}

For  $\Delta I=1/2$ transitions and $\epsilon^{\prime}/\epsilon$,
the subtractions of the power divergencies, necessary to obtain  finite 
matrix elements,   are the major obstacle in lattice calculations. 
These divergencies arise from penguin contractions of the relevant 
operators, fig.~\ref{fig:dp}, which induce a  mixing with operators of lower dimensions,  
namely $\bar s \sigma^{\mu\nu} G^{A}_{\mu\nu} t^{A} d$, $\bar s 
\sigma^{\mu\nu} \gamma_{5} G^{A}_{\mu\nu} t^{A} d$, $\bar s d$ and 
$\bar s \gamma_{5} d$~\cite{renorm}. Power divergencies and subtractions are 
encountered with both the strategies ($K\to \pi\pi$ or $K \to 
\pi$) adopted to compute the physical  amplitudes. 
For example, with a propagating charm quark,   $Re\, {\cal A}_{0}$ is
computed in terms of the matrix elements of $Q^{\pm}$ ($Q_{1}$ and 
$Q_{2}$)  only. 
In this case, due to the GIM mechanism, the subtraction is implicit  in the difference of 
penguin  diagrams with a charm and an up quark propagating in the 
loop:
\beq Q^{\pm} = \bar s \gamma_{\mu} \left(1-\gamma_{5} \right) 
u \, \bar u  \gamma^{\mu} \left(1-\gamma_{5} \right)  d \pm  \bar s \gamma_{\mu} \left(1-\gamma_{5} \right) 
d \, \bar u  \gamma^{\mu} \left(1-\gamma_{5} \right) u  - (u \to c)  \, .\eeq

In the past any attempt to compute $\Delta I=1/2$  $K \to \pi\pi$ matrix elements 
failed because no visible signal was observed after the subtraction of 
the power divergencies~\cite{sonicapri89,marticapri89}. These calculations, performed 
about twelve years ago with very modest computer resources, on small 
lattices and with  little statistics, were abandoned after the 
publication of the Maiani-Testa no-go theorem~\cite{mt}.  

This year, the SPQ$_{CD}$R Collaboration has found the 
first signals for both the matrix elements of $Q^{-}$ and $Q_{6}$, 
figs.~\ref{fig:qm}, \ref{fig:q6} and \ref{fig:qp}.  The 
difference with respect to previous attempts is given by a much larger volume and 
statistics, 
about 400 configurations  with $V=24^{3} \times 64$, compared to a 
few  tens of configurations on lattices with $V=16^{3} \times 32$ (8 
configurations on $V=24^{3} \times 
40$)~\cite{sonicapri89} or 110 configuration of a lattice with $V=16 
\times 12 \times 10 \times 32$ (sic !!)~\cite{marticapri89}. A crucial 
ingredient is also  to work  at the point corresponding to  the matching condition $m_{K} = 
W_{0}$, where $W_{0}$ is the energy of the two pions at rest on the 
finite volume used in the simulation, namely with non degenerate strange 
and light quark masses~\footnote{ 
In this case, as in the case of degenerate quark masses, the 
subtraction of operators of lower 
dimensions is not necessary~\cite{dawson}.}.
 Past calculations were always performed, instead, at the  degenerate point, $m_{K} = m_{\pi}$.   The 
results are very preliminary and in particular the two-pion energy in 
the finite volume for this channel gives a scattering lenght,  
$a_{I=0}$, in disagreement  with expectations  both in value and in the mass dependence.  This is 
to be contrasted with the $\Delta I=3/2$ case, where the analysis of 
the scattering length $a_{I=2}$ is in good agreement with 
expectations~\cite{spqr}. 
Much more work is needed to clarify  these problems  before trying to 
extrapolate the amplitude to the physical point.  

At the matching point $m_{K} = W_{0}$, there are also results 
for  $\langle \pi \pi \vert Q_{6}\vert K \rangle$. In 
this case there is no GIM mechanism at work, and a finite subtraction 
of the matrix element of the pseudoscalar density operator $Q_{P} = 
(m_{s} -m_{d}) \bar s \gamma_{5} d $ must be done (for parity, 
instead, the subtraction of the scalar operator $Q_{S} = (m_{s} + m_{d}) 
\bar s  d $ is not necessary). The  coefficient of the mixing  of $Q_{P}$ with $Q_{6}$ is quadratically divergent~\footnote{ The 
mixing with the chromomagnetic operator is not discussed here. This mixing 
is a small, finite correction  which can be ignored to simplify the  discussion.}
\beq \hat Q_{6} \sim Q_{6} + \frac{C_{P}(\alpha_{s})}{a^{2}} Q_{P} \eeq
If not for the  explicit chiral symmetry breaking of the lattice  
action,  $\langle \pi \pi \vert Q_{P}\vert  K\rangle$  would vanish by the equation of motions  when 
$m_{K} = W_{n^{*}}$. With 
an improved action, $\langle \pi \pi \vert Q_{P} \vert K\rangle$ is 
of ${\cal O}(a^{2})$   and thus a finite subtraction is necessary.  The 
preliminary results shown in fig.~\ref{fig:qp} show that that the value 
of   $\langle \pi \pi \vert \bar s \gamma_{5} d  \vert K\rangle$ drops by a factor of  about 20 with respect 
to  $\langle \pi \pi \vert Q_{6} \vert K\rangle$  
when $m_{K} = W_{0}$ (it is of the same order when the kinematics is not matched)  and thus the subtraction is not very critical. The 
signal itself is very noisy however, see fig.~\ref{fig:q6}, and even with 400 configurations 
the statistical error on  $\langle \pi \pi \vert Q_{6} \vert K\rangle$  is about  $50\, \%$. Thus 
 a rather large statistics will be  necessary to obtain a relatively accurate result. 

$K \to \pi \pi$ amplitudes in the chiral limit can also be 
extracted from the calculation of $K\to \pi$ matrix elements, and indeed  
this has been the most popular method in  lattice 
calculations of $\Delta I =3/2$ transitions.  The main advantage of  this approach is that the 
three-point correlators, necessary for the extraction of the matrix 
elements, are less noisy than the four-point correlators used in the  $K\to 
\pi\pi$ case. Moreover, for  $K\to \pi$ matrix elements,  finite volume corrections  
are exponentially small as $L \to \infty$.   The main disadvantage is 
that the $K \to \pi\pi$  amplitudes are obtained, using soft pion 
theorems,  in the chiral limit only  and then the 
effect of FSI is definitively lost.  If these are important for  $\Delta 
I=1/2$ channels there is no hope, then, to recover the physical 
amplitudes. 

\begin{figure}[htb!]
\begin{center}
{\epsfig{figure=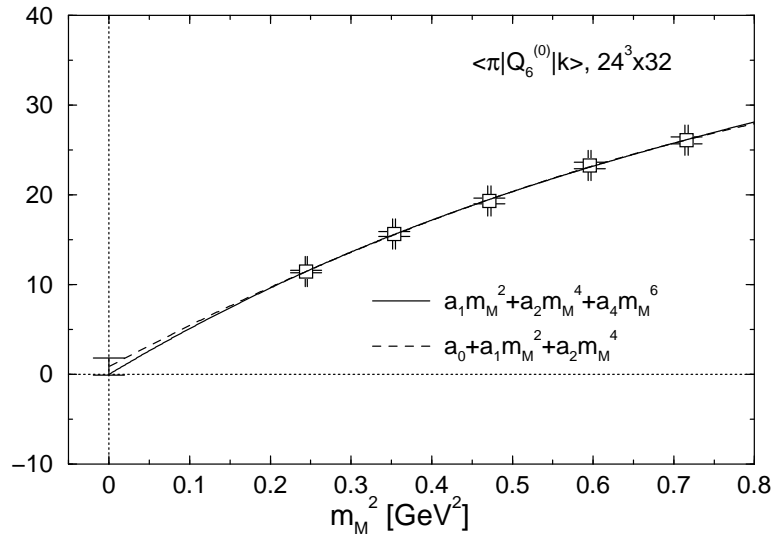, width=0.6\linewidth}}
\end{center}
\caption{{\it  Chiral behaviour of the subtracted matrix element $\langle 
\pi \vert Q_{i}^{sub} \vert K \rangle$ from the CP-PACS 
Collaboration. The curves represent different fits used to extrapolate 
to the chiral limit.}}
\label{fig:cppacschiral}
\end{figure}

In order to make the $K \to \pi$ matrix element finite  at least a subtraction 
is necessary,  even in the absence of explicit chiral symmetry breaking in the 
action.    With Wilson-like  Fermions, for which chiral symmetry is explicitly broken, 
the  number of subtractions  makes this approach unpracticable as 
also demonstrated  by the failure  of  past  
attempts~\cite{sonicapri89,marticapri89}. The  
method has acquired a renewed popularity, instead,  with 
recent formulations of the lattice theory, DWF or 
overlapping Fermions,  for which chiral symmetry breaking is absent 
or   strongly reduced in practice.     In this case, and under the 
hypothesis that chiral symmetry is under control, a finite amplitude 
can be obtained  by subtracting the scalar density amplitude with a 
suitable coefficient $C_{i}$ which depends on the operator at hand
\beq \langle \pi \vert Q_{i}^{sub} \vert K \rangle = \langle \pi \vert Q_{i} \vert K \rangle  - C_{i} \langle \pi \vert Q_{S}\vert K \rangle \, .
\label{eq:subkpi}  \eeq
The power divergent coefficients $C_{i}$ cannot be computed 
perturbatively.  Following  ref.~\cite{wise}, when  chiral symmetry is 
preserved,  the $C_{i}$ can be determined  by  the condition that the $K \to 0$   matrix element 
of the subtracted operator vanishes
\beq \langle 0 \vert Q_{i}  - C_{i} Q_{P} \vert K \rangle =0 \, . \eeq
The coefficients $C_{i}$ have been  obtained either  using  non degenerate 
quarks, $m_{s} \neq m_{d}$, by the RBC Collaboration,  or from the 
derivatives  of the 2-point correlation function with respect to the quark mass, 
by  the CP-PACS Collaboration.
\begin{figure}[htb!]
\begin{center}
{\epsfig{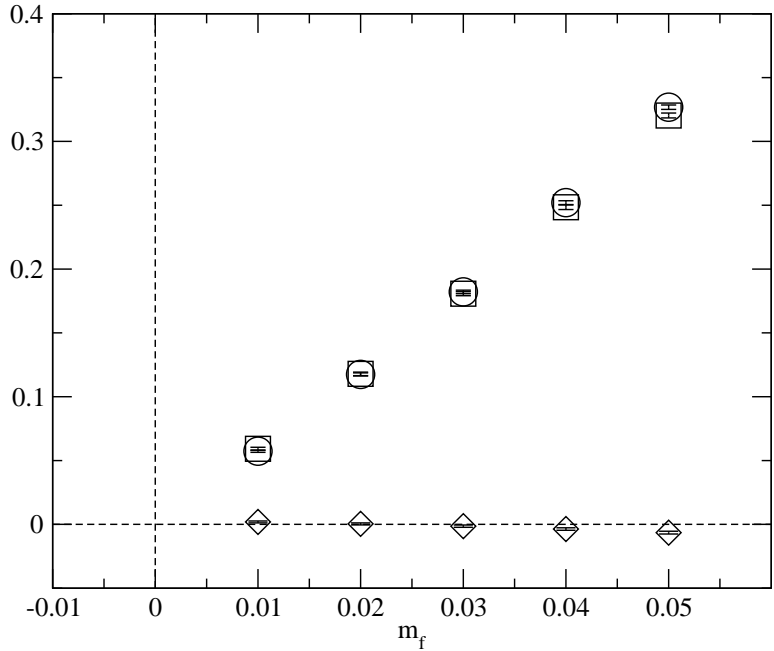}}
\end{center}
\caption{{\it   $\langle  \pi 
\vert Q \vert K \rangle$ for the unsubtracted operator,  $Q_{6}$ 
(squares),  the subtraction, $C_{6} Q_{S}$ (Circles), and the total, 
$Q_{i}^{sub}$ see eq.~(\ref{eq:subkpi}), as a function  of the quark mass 
$m_{f}$. The results are from the RBC Collaboration.}}
\label{fig:subrbc}
\end{figure}
Numerically, the  matrix element of the subtracted operator is much smaller than the 
unsubtracted one, as shown by  fig.~\ref{fig:subrbc}, taken from the 
recent work of the RBC collaboration~\cite{rbc2001}. This implies that any systematic uncertainty which 
enters in the subtraction procedure can have huge effects in the 
determination  of the physical amplitudes. After the subtraction, the 
chiral dependence of the matrix element has to be fitted in order to 
extrapolate it to the chiral limit. Both groups have included the 
logarithmic corrections which arise in quenched $\chi$PT in the fit 
and, in some cases, polynomial corrections of higher order in 
$m^{2}_{\pi}$. The chiral behaviour observed by  CP-PACS  is very 
satisfactory, as shown by fig.~\ref{fig:cppacschiral}, less good in the 
case of the RBC Collaboration. The difference may be due to the fact 
that a different gluon action is used in the two cases,  corresponding 
to smaller chirally breaking effects for CP-PACS. This can be monitored by 
measuring the so called ``residual mass'' which should be zero for 
perfect chirality and  is about a factor of ten smaller for CP-PCAS 
than for RBC. For a more extended discussion on this important point 
the reader can refer to the forthcoming proceedings of the Lattice 2001 
Conference~\cite{latt2001}, or those of the 2000 
edition~\cite{latt2000}.  

Let me also mention that the overall 
renormalization constants have been computed perturbatively by 
CP-PACS~\cite{aoki} and non-perturbatively by 
RBC~\cite{rbcnp}.
My compendium  of the physics results obtained by the two groups is given 
in table~\ref{tab:rbcpacs}
 \begin{table}[htb]
\centering
\caption{{\it Lattice results for $\Delta I=1/2$ transitions using $K \to \pi$ 
matrix elements from RBC and  CP-PACS. The experimental numbers are 
also given.}}
\label{tab:rbcpacs}
\begin{tabular}{||c|c|c|c|c||}\hline\hline
Reference & $Re {\cal A}_{0}$ & $Re {\cal A}_{2}$ & $Re {\cal A}_{0}/Re {\cal A}_{2}$   & $\epsilon^{\prime} / \epsilon$ \\
\hline \hline
RBC~\cite{rbc2001} 2001 & $29 \div 31 \times 10^{-8}$ & $1.1 \div 1.2 \times 
10^{-8}$ & $24 \div 27$  &  $-8 \div -4 \times 10^{-4}$ \\
CP-PACS~\cite{cppacs2001} 2001 & $16 \div 21 \times 10^{-8}$ & $1.3 \div 1.5 \times 10^{-8}$ & 
$9 \div 12$ &  $-7 \div -2 \times 10^{-4}$\\
Exps.~\cite{na48,ktev,PDG} 2001& $33.3 \times 10^{-8}$ &    $1.5 \times 
10^{-8}$ &  $ 22.2 $ & $17.2 \pm 1.8  \times 10^{-4}$ \\
\hline \hline
\end{tabular}
\end{table}

A few observations are necessary at this point. First of all,  $Re {\cal 
A}_{0}$, and consequently $Re {\cal 
A}_{0}/Re {\cal A}_{2}$, from the RBC  Collaboration are in good agreement with 
experimental values, contrary to CP-PACS which finds $Re {\cal 
A}_{0}$ smaller   by about a factor of two than the experimental number. 
Since the two groups use the same lattice formulation of the theory 
and differ only by, hopefully, marginal details, the reason of this 
difference should be clarified.  In both cases, however,  $\epsilon^{\prime} / 
\epsilon$ is in total disagreement with the data. The main reason is 
that the value of the matrix element  of $Q_{6}$ is much smaller than 
what would be necessary to reproduce the experimental value  (it corresponds 
approximatively to $B_{6} = 0.3\div 
0.4$), see also the talk  by J. Donoghue at  this Conference.  
Let me list a number of sources of systematic errors which may explain 
these embarassing results:
\begin{enumerate} 
\item \underline{Chirality} By working with DWF at a finite fifth 
dimension, $N_{5} =16$,  
a residual chiral symmetry  breaking is present in the theory. The 
amount of residual symmetry breaking is parametrized by a mass scale 
denoted by $m_{res}\sim 0.2 \div 2.0$~MeV. In the 
presence of explicit chiral symmetry breaking, the coefficients 
$C_{i}$ determined from $K \to 0$  differs from the correct one and 
this may induce an  error of ${\cal O}(m_{res} a^{-2}) \sim 
(200 \, 
\mbox{MeV})^{3}$ on matrix elements which are of the order of 
$\Lambda_{QCD}^{3} \sim (300  \, \mbox{MeV})^{3}$.   Both groups 
claim to have this point fully under control~\cite{cppacs2001,rbc2001}. 
A calculation at a larger value of $N_{5}$, with all the other 
parameters unchanged, would be very useful to clarify the situation. 
\item \underline{Matching below the charm mass} Both groups have so far presented results at a renormalization 
scale just  below the charm quark mass. As discussed in the 
introduction, the  matching of the effective theory is rather problematic at such low scales.
\item \underline{Extra Quenched Chiral Logarithms} As 
shown by fig.~\ref{fig:subrbc}, the subtraction of the power divergencies is very critical. Besides the effects discussed in 1., 
there is another delicate problem  which has been  recently rised by 
Golterman and Pallante~\cite{goltepallalast}. In the quenched theory  the 
$(8,1)$ operators, such as $Q^-$ and $Q_{6}$, do not belong anymore to 
irreducible representations of the chiral group and this gives rise to spurious chiral logarithms. These logarithms affect the subtraction 
procedure and   have not been taken into account in  the analyses of RBC and CP-PACS.
\item \underline{Higher order chiral corrections and FSI} Higher 
order chiral contributions, among which FSI have also to be accounted, 
can produce  large variations  of the matrix elements  between the 
chiral limit and the physical point.  In simulations where only  $K \to \pi$ 
matrix elements are computed,  $K \to \pi\pi$  amplitudes can only be 
obtained at lowest order in $\chi$PT where these physical effects are 
missing.
\item \underline{New Physics} As noticed by Murayama and 
Masiero~\cite{murayama}, it 
is possible to produce large effects for $\epsilon^{\prime} /\epsilon$ 
in SUSY without violating other bounds coming from FCNC. If the 
lattice results are correct, this is an open possibility. Before 
invoking new physics,   at least the result for $Re{\cal 
A}_{0}$ should be established with more confidence. Given that the two 
groups still don't agree on this quantity, more work is needed.
\end{enumerate}
\section{Conclusion and outlook}
A renewed activity in lattice calculations of kaon decays and mixing  
has developed in the last two years.
For $\Delta I=3/2$ transitions, $K^{+} \to \pi^{+} \pi^{0}$ and 
electroweak penguin amplitudes, new and more  precise results have 
been obtained. In this case, removal of lattice artefacts, by 
extrapolation to the continuum and/or improvement, and unquenched 
calculations are around the corner.
For $\Delta I=1/2$ transitions,  it has been shown that direct computations 
of  $K \to \pi\pi$ amplitudes, including FSI,  is 
possible~\cite{lll,lmst}, although 
the pratical implementation with sufficient accuracy will require more 
time.  First results, obtained by  using $K \to \pi$ lattice matrix elements 
computed with DWF and soft-pion theorems,  show a striking 
disagreement with the experiments for $\epsilon^{\prime}/\epsilon$.  
For $Re {\cal A}_{0}$,  there is  a factor of two between the values 
found by the CP-PACS~\cite{cppacs2001}  and 
RBC~\cite{rbc2001} Collaborations. More work is needed in this case to clarify all 
these points. Many related physical quantities, like the strong 
interaction  phase-shifts for $\pi \pi$ scattering, the 
chromomagnetic operator matrix elements,  semileptonic $K \to \pi\pi 
l \nu_{l}$ amplitudes and the scalar semileptonic form factor will also be 
obtained as a byproduct from  lattice investigations of non-leptonic kaon decays, see also J.~Donoghue 
at this Conference.  
\section*{Acknowledgements}
I thank  particularly D.~Becirevic, M.~Golterman R.~Gupta, D.~Lin, 
V.~Lubicz, R.~Mawhinney,
J.~Noaki, M.~Papinutto and  S.~Sharpe for  fruitful and   constructive   discussions. 
I am also particularly indebted to all the members of the SPQ$_{CD}$R 
Collaboration.  This work was supported by European Union grant HTRN-CT-2000-00145.

\end{document}